\documentclass{PoS}

\title{Novel Heavy Quark Phenomena in QCD}

\ShortTitle{Novel Heavy Quark Phenomena in QCD}

\author{\speaker{Stanley J. Brodsky}
\\
        SLAC National Accelerator Laboratory\\
        E-mail: \email{sjbth@slac.stanford.edu}}

\abstract {Heavy quarks provide  a new dimension to QCD, allowing tests of fundamental  theory, the nature of color confinement, and the production of new exotic multiquark states.   I also discuss novel explanations for several apparently anomalous experimental results, such as the large $t \bar t$ forward-backward asymmetry observed in $p \bar p$ colisions at the Tevatron,  the large rates for $\gamma$ or $Z$ plus high-$p_T$ charm jets observed at the Tevatron, the strong nuclear absorption of the $J/\psi$ observed in $pA$ collisions at the LHC, as well as fixed target experiments at high $x_F$.  Precision measurements of the heavy quark distribution in hadrons at high $x$ are needed since intrinsic heavy quarks can play an important role in high $x$ phenomenology as well as predicting a new mechanism for high-$x_F$ Higgs production. The  role of  multi-parton interactions, such as di-gluon initiated subprocesses for forward quarkonium hadroproduction, is discussed. I also briefly discuss a new approach to the QCD confinement potential and the origin of the QCD mass scale based on AdS/QCD, light-front holography and a unique extension of conformal theory.  The renormalization scale ambiguity can be eliminated at finite orders in pQCD using the scheme-independent PMC procedure, thus increasing the precision of predictions and eliminating an unnecessary source of theoretical systematic error.}

\FullConference{XV International Conference on Hadron Spectroscopy-Hadron 2013\\
		4-8 November 2013\\
		Nara, Japan }

\begin{document}

\section{Introduction}

Quantum Chromodynamics is a truly remarkable theory, describing the spectroscopy and dynamics of hadron and nuclear physics in terms of fundamental, but confined, quark and gluonic non-Abelian gauge  fields.  The existence of heavy quarks gives a new dimension to QCD, allowing tests of the theory and the nature of color confinement in many diverse ways.  The discovery of the charged $Z^+$ states~\cite{Ablikim:2013emm},  which can be interpreted as  a $c \bar c u \bar d$ heavy tetraquark state~\cite{Matheus:2006xi,Karliner:2013dqa},  is just a hint of the vast array of exotic color-singlet resonances or bound states of light and heavy quarks possible in QCD.   

The very strong spin correlation~\cite{Court:1986dh} $A_{NN}$ appearing at  $\sqrt s \simeq~ 3 ~{\rm and} ~5 ~{\rm GeV}$ in large-angle elastic proton-proton scattering  $pp \to pp$ can be understood as due to the excitation of  charge $ Q=+2$ baryon number $B=2$ $ |uud uud Q \bar Q \rangle$ resonances  in the $s$-channel at the strangeness and charm thresholds~\cite{Brodsky:1987xw}.   
Similarly, one expects color-singlet $Q=+4$ hexaquark resonances~\cite{Bashkanov:2013cla} $|uuu uuu\rangle$, $|uuu uuc\rangle,$   and $uuu ucc\rangle$ in which the six $3_C$ quarks are bound as a color-singlet S-wave configuration,    analogous to the  ``hidden-color" configurations~\cite{Brodsky:1983vf,Harvey:1980rva,Brodsky:1976mn} which dominate the dynamics of the deuteron at short distances.
In addition, the attractive multi-gluon exchange van der Waals potential leads to the prediction~\cite{Brodsky:1989jd} of  ``nuclear-bound quarkonium" states such as $[Q \bar Q  A].$

Hadrons in QCD are eigenstates of the light-front Hamiltonian $H^{QCD}_{LF} |\Psi_H\rangle = M^2_H  |\Psi_{LF}\rangle$, the evolution operator in light-front time $\tau = t +z/c$.  The hadronic eigenfunction can be projected on the $n$-particle eigensolutions $| n \rangle$ of the free QCD Hamiltonian to generate the frame-independent LF Fock state wavefunctions 
$\psi^H_n(x_i, \vec k_{\perp i}, S^z_i ) = \langle\Psi_H | n\rangle$, where $x_i= {k^0+k^z \over P^0 +P^z}={k^+_i \over P^+}$  
(with  $\sum_i^n x_i =1$)  are the quark and gluon light-front momentum fractions, the $\vec k_{\perp i}$  (with $\sum^n_i \vec k_{\perp i} = 0_\perp$) are the transverse momenta,
and the $ S^z_i $ are  the constituent spin-projections in the $\hat z$ direction which satisfy $J^z$ angular momentum conservation. Given its LF wavefunction, one can compute the quark and gluon composition of a hadron  and thus virtually all its hadronic observables.  For example, the square of the LFWFs generate structure functions and their overlaps determine the elastic and inelastic form factors.  For a review see ref. \cite{Brodsky:1997de}.  The entire hadronic spectrum in  QCD (1+1), including higher Fock states, can be systematically computed using the discretized light-cone quantization (DLCQ) method~\cite{Pauli:1985pv}

In general, the hadron eigensolution has an infinite number  of distinct Fock states;  higher Fock states such as  $|uud Q \bar Q\rangle$ generate the heavy sea quark distributions, both the ``extrinsic" contributions corresponding to gluon splitting  $g \to Q \bar Q$ predicted by DGLAP evolution, plus the ``intrinsic" contributions~\cite{Brodsky:1980pb} in which the heavy quarks are multi-connected to the valence quarks.  One can show from the operator product expansion~\cite{Brodsky:1984nx,Franz:2000ee}   that the probability of intrinsic heavy quarks  in a light hadron decreases as $1/ M^2_Q$, corresponding to the twist-6 operator $G^3_{\mu \nu}$ in non-Abelian QCD.   In contrast, the fall-off in Abelian QED is $1/ M^4_Q$, corresponding to the twist-8 Euler-Heisenberg light-by-light scattering operator $F^4_{\mu \nu}$. 
  
\section{The Unique Color-Confining Potential}

The QCD Lagrangian has no explicit mass scale if all quark masses are set to zero.  The classical QCD Lagrangian and its action thus have a fundamental conformal symmetry in the limit of massless quarks.  Remarkably, as first shown by de Alfaro, Fubini, and Furlan~\cite{de Alfaro:1976je}, the action will retain its conformal invariance even if the Hamiltonian is augmented by mass terms proportional to the dilatation operator $D$ and the special conformal operator $K$.  If one applies the dAFF formalism to light-front Hamiltonian theory, the $q\bar q$ interaction has a unique form of a confining harmonic oscillator potential~\cite{Brodsky:2013ar} which depends on a single mass-scale parameter $\kappa$.    The result is a nonperturbative relativistic light-front quantum mechanical wave equation for mesons which incorporates color confinement, a mass gap, and other essential spectroscopic and dynamical features of hadron physics. 

The same light-front equations arise from the holographic mapping of the soft-wall model modification of AdS$_5$ space -- with a unique dilaton profile $e^{\kappa^2 z^2}$ -- to QCD (3+1) at fixed light-front time $\tau$.  ``Light-front holography"~\cite{deTeramond:2008ht} thus provides a precise relation between amplitudes in the fifth dimension of AdS space and light-front wavefunctions.

The effective $q \bar q$ color-confining interaction between light quarks in the LF Hamiltonian derived from AdS/QCD and light-front holography has the unique form of a two-dimension harmonic oscillator $U(\zeta^2)  = \kappa^4 \zeta^2 + 2\kappa^2(J-1)$  where the invariant variable $\zeta^2= b^2_\perp x(1-x)$  is conjugate to the invariant mass squared.  A complimentary argument for the form of the LF-confining potential is given in ref.~\cite{Glazek:2013jba}.
The eigensolutions of this ``Light-Front Schr\"odinger Equation" correspond to a massless pion for zero quark mass and linear Regge trajectories with the same slope in the radial quantum number $n$ and orbital angular momentum $L$~\cite{Brodsky:2013dca}.   The resulting LF wavefunctions have remarkable phenomenological features.  For example, the proton eigensolution of the corresponding LF Dirac equation has equal probability to have relative quark-diquark relative angular momentum $L=0$ and $L=1$.  Forshaw and Sandapen~\cite{Forshaw:2012im} have shown that the resulting predictions for $\rho$ electroproduction agree with experiment. 
The quark counting rules~\cite{Brodsky:1973kr,Matveev:1973ra} for hard exclusive processes are first-principle features of AdS/QCD~\cite{Polchinski:2001tt}.   The shape of the QCD running coupling in the nonperturbative domain is also predicted~\cite{Brodsky:2010ur}, in agreement with effective charge phenomenology for 
 $Q^2 < 1~{\rm GeV}^2$.
The AdS/QCD light front approach has recently been extended to heavy quarks, successfully  describing the spectroscopy, wavefunctions and decays of heavy hadrons~ \cite{Ahmady:2013cga,Forshaw:2012im,Branz:2010ub}.

\section{Anomalous Heavy-Quark Measurements}

A number of recent experimental results involving heavy quarks  appear to be in striking disagreement with conventional expectations.

\subsection{Top/Anti-Top Asymmetry and PMC Renormalization Scale Setting}

The $t$ versus $\bar t$ momentum asymmetry measured at the Tevatron  by CDF ~\cite{Aaltonen:2011kc}  and by D0~\cite{Abazov:2011rq} in $\bar p p \to t \bar t X$ disagrees with canonical PQCD predictions by more than $3 \sigma.$
However, as Xing-Gang Wu and I have shown in refs. ~\cite{Brodsky:2011ig,Brodsky:2012ik}, this disagreement can be attributed to an arbitrary, scheme-dependent choice of the renormalization scale of the QCD running coupling constant $\alpha_s(\mu^2)$.  In contrast, when one uses the scheme-independent
Principle of Maximum Conformality (PMC)~\cite{Brodsky:2011ig,Brodsky:2013vpa} to set the scale, the discrepancy between pQCD prediction and experiment is reduced to 1 $\sigma. $ 

The  running
coupling in a gauge theory  sums  the terms involving the
$\beta$ function; thus when the renormalization scale is set
properly, all non-conformal $\beta \ne 0$ terms  in a perturbative
expansion arising from renormalization are summed into the running
coupling. 
The remaining terms in the perturbative series will then be
identical to those of a conformal theory; i.e., the corresponding
theory with $\beta=0$. 
As discussed by Di Giustino, Wu, Mojaza, and myself~\cite{Brodsky:2011ig,Brodsky:2011ta,Mojaza:2012mf}, the resulting scale-fixed predictions using
the PMC are independent of
the choice of renormalization scheme --  as required  by the 
renormalization group.    The PMC is the principle~\cite{Brodsky:2011ig,Brodsky:2011ta}
which underlies the BLM scale-setting 
method.~\cite{Brodsky:1982gc} 

The PMC/BLM scales are fixed order-by-order and the 
scales then automatically determine the number $n_f$ of effective flavors in the $\beta$-function analytically~\cite{Brodsky:1998mf}. The results avoid the divergent renormalon
resummation~\cite{Brodsky:2000cr} and agree with QED scale-setting in the Abelian limit.  In the case of QED, the PMC scale is proportional to the photon virtuality and  thus sums all vacuum polarization corrections to all orders. 
Different schemes lead to different effective PMC/BLM scales, but the final results are scheme independent. The PMC procedure is also valid for multi-scale processes. 

One can introduce a generalization of conventional dimensional regularization, 
the ${\cal R}_\delta$ schemes. For example, if one generalizes the $\bar{MS} $ scheme by subtracting $\ln 4 \pi - \gamma_E - \delta$  instead of just $\ln 4 \pi - \gamma_E$  
the new terms generated in the pQCD series that are proportional to $\delta$ expose the $\beta$ terms and thus the renormalization scheme dependence.
Thus the   ${\cal R}_\delta$ schemes uncover the renormalization scheme and scale ambiguities of pQCD predictions, expose the general pattern of nonconformal terms, and allow one to systematically determine the argument of the running coupling order-by-order in pQCD in a form which can be readily automatized~\cite{Mojaza:2012mf, Brodsky:2013vpa}.
The resulting PMC scales and  the finite-order PMC predictions are to high accuracy independent of the choice of the initial renormalization scale.  

The PMC satisfies  all of the principles of the renormalization group: reflectivity, symmetry, and transitivity, and it thus eliminates an unnecessary source of systematic error in pQCD predictions~\cite{Wu:2013ei}.  The BLM/PMC also provides scale-fixed,
scheme-independent high-precision connections between observables, such as the ``Generalized Crewther Relation''~\cite{Brodsky:1995tb}, as well as other ``Commensurate Scale Relations''~\cite{Brodsky:1994eh,Brodsky:2000cr}.

The renormalizations scales for multi-scale amplitudes are also determined by the PMC. For example, the PMC/BLM scale~\cite{Brodsky:1995ds} for the running coupling  appearing in the final state for heavy-quark production at threshold is proportional to the relative velocity 
within the $Q \bar Q$ pair, very different than the usual assumption that the renormalization scale is of order of the heavy quark mass.  

The elimination
of the renormalization scheme ambiguity 
thus improves the accuracy  of pQCD tests and 
increases  the sensitivity of LHC experiments and other measurements to new physics
beyond the Standard Model.

\subsection{High-$x$ Strangeness Distributions}

\begin{figure}
 \begin{center}
\includegraphics[height=10cm,width=15cm]{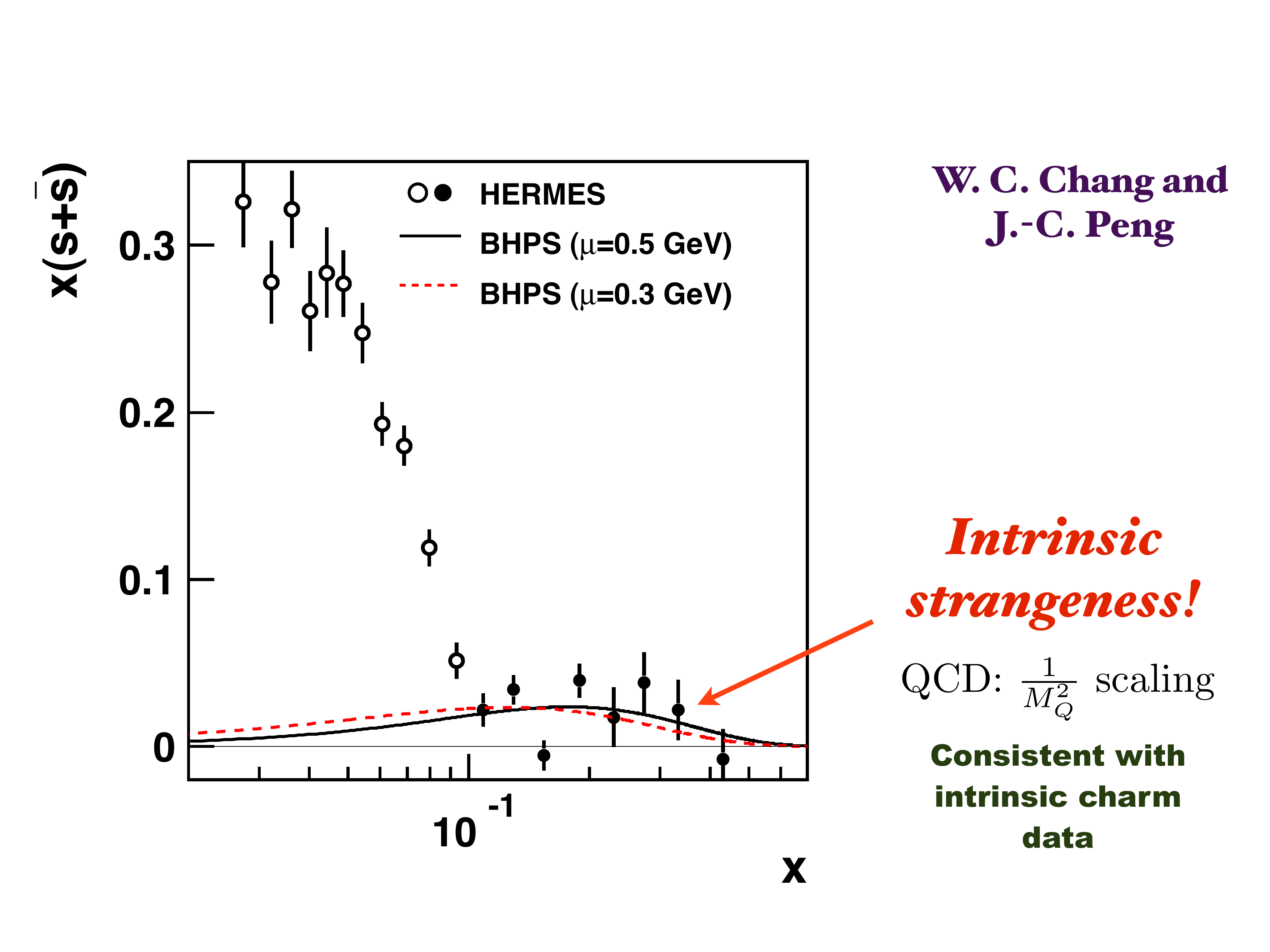}
\end{center}
\caption{Hermes measurement of the strangeness distribution compared with the BHPS model. From Chang and Peng~\cite{Chang:2011du} 
}
\label{Hermes}  
\end{figure}

The strange quark $s(x,Q^2) + \bar s(x,Q^2) $ distribution in the proton measured by HERMES in $\gamma^*  p \to K X$ reactions~\cite{Airapetian:2008qf}  has significant support  at large $x_{bj}$, in contradiction with usual expectations.  The strange quark in the proton is seen to have two distinct components: a fast-falling contribution,  consistent with gluon splitting to $s \bar s$, and an approximately flat component up to $ x < 0.5$.  See fig. \ref{Hermes}.   As emphasized by Chang and Peng~\cite{Chang:2011du}, 
the ``intrinsic" component~\cite{Brodsky:1980pb} at high $x$ is in agreement with expectations derived from the nonperturbative 5-quark light-front (LF) Fock state $|uuds\bar s\rangle$  of the proton. The dominant configuration in $x_i$ and $k_{\perp i}$ will minimizes the 
total invariant mass: $ {\cal M}^2= \sum_{i=1}^5 {k^2_{\perp i } +m_i^2\over x_i}$; i.e. minimal rapidity differences of the constituents.  Equal rapidity implies that the light-front momentum fractions are proportional to the quark transverse mass: $x_i \propto m_{\perp i}$ 
with $m_{\perp i} = \sqrt {k^2_{\perp i} + m^2_i}. $  The heavy quarks in the$|uudQ\bar Q\rangle$  Fock state thus have the largest LF momentum fraction $x_i$.

The high-$x$ intrinsic strange quarks can reinteract with the valence quarks in the $|uud s \bar s\rangle$ Fock state since all of the constituents in the LF Fock state tend to have the same rapidity. 
This leads to a  $s(x,Q^2)$ versus  $ \bar s(x,Q^2) $ asymmetry in both momentum and spin, as also expected when one identifies~\cite{Brodsky:1996hc} the $|uud s \bar s\rangle$ Fock state with the analogous $ | K^+(\bar s u) \Lambda(sud)\rangle$ hadronic state.  Similarly, the  $\bar u(x) \ne \bar d(x)$ asymmetry can be identified with the nonperturbative dynamics of the $|uud q \bar q\rangle$ Fock state.

\subsection{High Transverse Momentum Heavy Quark Jet Production and the Physics Consequences of Intrinsic Heavy Quarks }

The cross sections for  high transverse photon plus a charm jet cross section $ \bar p  p \to \gamma c X$ and also $Z^0$ plus a charm jet $ \bar p p  \to Z^0 c X$  measured at the Tevatron~\cite{Abazov:2009de} for  $p_T^\gamma > 60~ GeV/c$ appear to be substantially larger than predicted using conventional charm PDF distributions.   
In contrast, the corresponding rate for $\bar p p \to b + \gamma  X$ agrees  well with NLO PQCD predictions.

The dominant underlying 2 to 2 subprocesses~\cite{Stavreva:2010mw} are $g c \to \gamma  c $ and $g c \to Z^0 c$, which depend critically on the assumed parametrization of the charm quark PDFs at  $x>0.1$ and high $Q^2 \sim 10^4~{\rm GeV}^2$.    The charm distribution in the proton predicted by QCD includes an intrinsic component  -- the five-quark Fock state $|uud c \bar c\rangle$  derived from multigluonic couplings of the $c \bar c$ pair to the proton's valence quarks.   The intrinsic charm quarks appear at large $x$ since this minimizes the off-shellness of the LFWF.  In fact, the EMC determination~\cite{Aubert:1981ix} of $c(x,Q^2)$ at $x=0.42$ and $Q^2 = 75~{\rm GeV}^2$  is approximately 30 times larger than predicted by the soft distribution from  gluon splitting $g \to c \bar c$.  CTEQ parametrizations~\cite{Pumplin:2007wg} include the intrinsic charm as measured by EMC.  The photon plus charm-jet anomaly could possibly  be explained if one allows for a substantial intrinsic contribution to the charm structure function in $g c \to c \gamma$  at $Q^2 \sim 10^4~{\rm GeV}^2$, but one requires a factor of two increase in strength compared to the CTEQ PDF.  The reduction of the charm distribution at large $x$ due to DGLAP evolution is likely to have been  overestimated because one conventionally takes $m_c=0$: it is clearly important to evolve $c(x,Q^2)$ to the high $Q^2 \sim 10^4~{\rm GeV}^2$ domain using massive charm quark in the DGLAP evolution equations;  The argument of the running coupling and the effective number  of flavors in the QCD $\beta$ function in the DGLAP evolution equations can be set using the scheme-independent  PMC method. These questions are now being investigated by Gang Li and myself.  The ratio of intrinsic charm to intrinsic bottom scales 
as $m_c^2/m_b^2 \simeq 1/10,$  using the operator product expansion in non-Abelian QCD,~\cite{Brodsky:1984nx,Franz:2000ee} so that intrinsic bottom plays a minor role in the Tevatron measurements.

In the case of a hadronic high energy proton collision, such as $pp \to \Lambda_c X$  the high-$x$ intrinsic charm quark in  the proton's $|uud c \bar c\rangle$ Fock state can   coalesce with the co-moving $ud$ valence quarks in a projectile proton to produce a forward   $\Lambda_c(cud)$ baryon at the combined high momentum fraction $x_F = x_u + x_d + x_c$.   
Similarly,  the coalescence of comoving $b$ and $\bar u$ quarks from the $|uud \bar b b\rangle$  intrinsic bottom Fock state can explain the high $x_F$ production of the $\Lambda_b(udb)$,  which was first observed at the ISR collider at CERN~\cite{Bari:1991ty} in association with a positron from the decay of the associated high-$x_F$ B meson.   A similar mechanism predicts quarkonium hadroproduction at high $x_F$~\cite{Vogt:1994zf}.

The NA3 experiment~\cite{Badier:1982ae} has observed the hadroproduction of {\bf two} $J/\psi$s at high $x_F$, a signal for seven-quark Fock states such as $|uud c \bar c c\bar c\rangle$  in the proton~\cite{Vogt:1995tf}. The intrinsic contributions can explain both the  open-charm and open-bottom hadron production at high momentum fractions, and it can also account for single and double $J/\psi$ hadroproduction  measured by NA3 at high $x_F$~\cite{Vogt:1992ki}.

Measurements by the SELEX collaboration~\cite{Engelfried:2005kd}  have led to the discovery  of a set of  doubly-charmed spin-1/2 and spin-3/2 baryons with quantum numbers that can be identified as 
$|ccu \rangle$ and   $|ccd \rangle $ bound states -- again a signal for seven-quark Fock states such as $|uud c \bar c c\bar c\rangle$  in the proton.  Surprisingly, the mass splittings of the $ccu$ and $ccd$ states measured by SELEX  are much larger than expected from known QCD isospin-breaking effects~\cite{Broadsky:2012rw}.
One  speculative proposal~\cite{Brodsky:2011zs} is that these doubly charmed baryons have the configuration $c~q~c$ where the light quark $q$ is exchanged between the heavy quarks, as in a linear molecule.  This  configuration may enhance the Coulomb repulsion of the $c~u~c$ relative to $c~d~c$.   It is clearly important to have experimental confirmation of the SELEX results.

The presence of intrinsic heavy quarks  in the proton leads to a novel mechanism for the inclusive and diffractive
Higgs production $pp \to p p H$ where  the Higgs boson carries a large fraction of the projectile proton momentum.~\cite{Brodsky:2006wb,Brodsky:2007yz}  This high $x_F$ production
mechanism is based on the subprocess $(Q \bar Q) g \to H $ where the Higgs couples to the sum of the momentum of the $Q \bar Q$ pair in the $\vert uud Q \bar Q \rangle$ intrinsic heavy quark Fock state of the colliding proton; it thus can be produced with approximately
$80\%$ of the projectile proton's momentum.   High-$x_F$ Higgs production could be measured at the LHC using far-forward detectors or arranging the proton beams to collide at a significant crossing angle or at different beam energies.  The same mechanism can produce the Higgs at large $x_F$ in $\gamma p \to X H$ collisions at the LHeC.  

Intrinsic charm in light hadrons also provides a solution to the $J/\psi \to \rho \pi$ puzzle~\cite{Brodsky:1997fj}.  The conventional  assumption is that quarkonium states decay into light hadrons via annihilation into virtual gluons, the OZI rule.   However, this hypothesis leads to the identical decays of the $J/\psi$ and the $\Psi^\prime$, up to a factor of $12\%$ from the wavefunction at the origin squared -- in strong disagreement with measurement.   Worse, the decay $J/\psi \to \rho \pi$ is predicted to be suppressed  by hadron-helicity conservation~\cite{Brodsky:1981kj}, when in fact it is the largest two-body hadronic decay. 
The $J/\psi \to \rho \pi$ puzzle can be explained if the $c \bar c$ does not annihilate but instead flows into the intrinsic charm Fock state $|q \bar q c \bar c\rangle$  of  one of the final state meson.  
In contrast, the  $\psi^\prime \to \rho \pi$ decay is suppressed by change of sign in the decay amplitude from the node in the $\psi^\prime$ wavefunction. 

Intrinsic charm also affects B-decays in a novel way~\cite{Brodsky:2001yt}. The presence of intrinsic charm in the hadronic light-front wave function of the $B$, even at a few percent level, provides new, competitive decay mechanisms for B decays which are nominally CKM suppressed.  The impact of intrinsic heavy quark distributions in the proton on new physics searches at the high intensity frontier is discussed in ref.~\cite{Brodsky:2012zza}.

Precision measurements of the charm  and bottom distributions in hadrons are clearly of prime interest.  This can be done in lepton-scattering facilities such as COMPASS, the LHeC, and the new lepton-ion colliders proposed at  BNL and JLab.  The proposed fixed target program ``AFTER" ~\cite{Brodsky:2012vg,Lansberg:2012kf} at the LHC will allow remarkable accessibility to heavy-quark phenomena in the target and projectile fragmentation regions of rapidity.   For example,  heavy charm and bottom hadrons  will be produced with relatively low rapidities in the easily instrumented target rapidity domain at AFTER.  Just as important, the existence of charm quarks at high momentum fraction in the proton wavefunction implies enhanced production of open and hidden charm states in the threshold regime.  Since the produced quarks and gluons are produced at threshold at small relative rapidity differences, this provides  an important  opportunity~\cite{Brodsky:2012zz} to create exotic heavy quark states at upcoming facilities such as JLab at 12 GeV, PANDA, and NICA.

\section{Nuclear Suppression of Quarkonium and Di-Gluon Saturation}

Since its radius is small, the cross section for the interaction of a charmonium state in nuclear matter is expected to be only a 
few millibarns, as expected from QCD color transparency~\cite{CT}.  Thus the usual expectation is that hadroproduction cross sections are  approximately linear in the number of nucleons $A$.   However, the production cross section $ p A \to J/\psi X$ measured by LHcB~\cite{Aaij:2013zxa} and ALICE~\cite{Aamodt:2010jd} at forward rapidity $y \sim 4$  shows an unexpectedly strong nuclear  suppression, close to $A^{2/3}$.    This effect cannot be accounted for by shadowing of the nuclear gluon distribution.  

Arleo and Peigne~\cite{Arleo:2012hn,Arleo:2012rs}  suggest that the strong nuclear suppression of $J/\psi$ production in $pA$ collisions can be explained as a manifestation of the ``color-octet" model:  the $c \bar c$ propagates through the nucleus as a color-octet, and its nuclear energy loss will be proportional to its energy if  the induced gluon radiation is coherent on the entire nucleus.  The color-octet $c \bar c$ is assumed to convert to the color-singlet  $J/\psi$ after exciting the nucleus.  

\begin{figure}
 \begin{center}
\includegraphics[height=10cm,width=15cm]{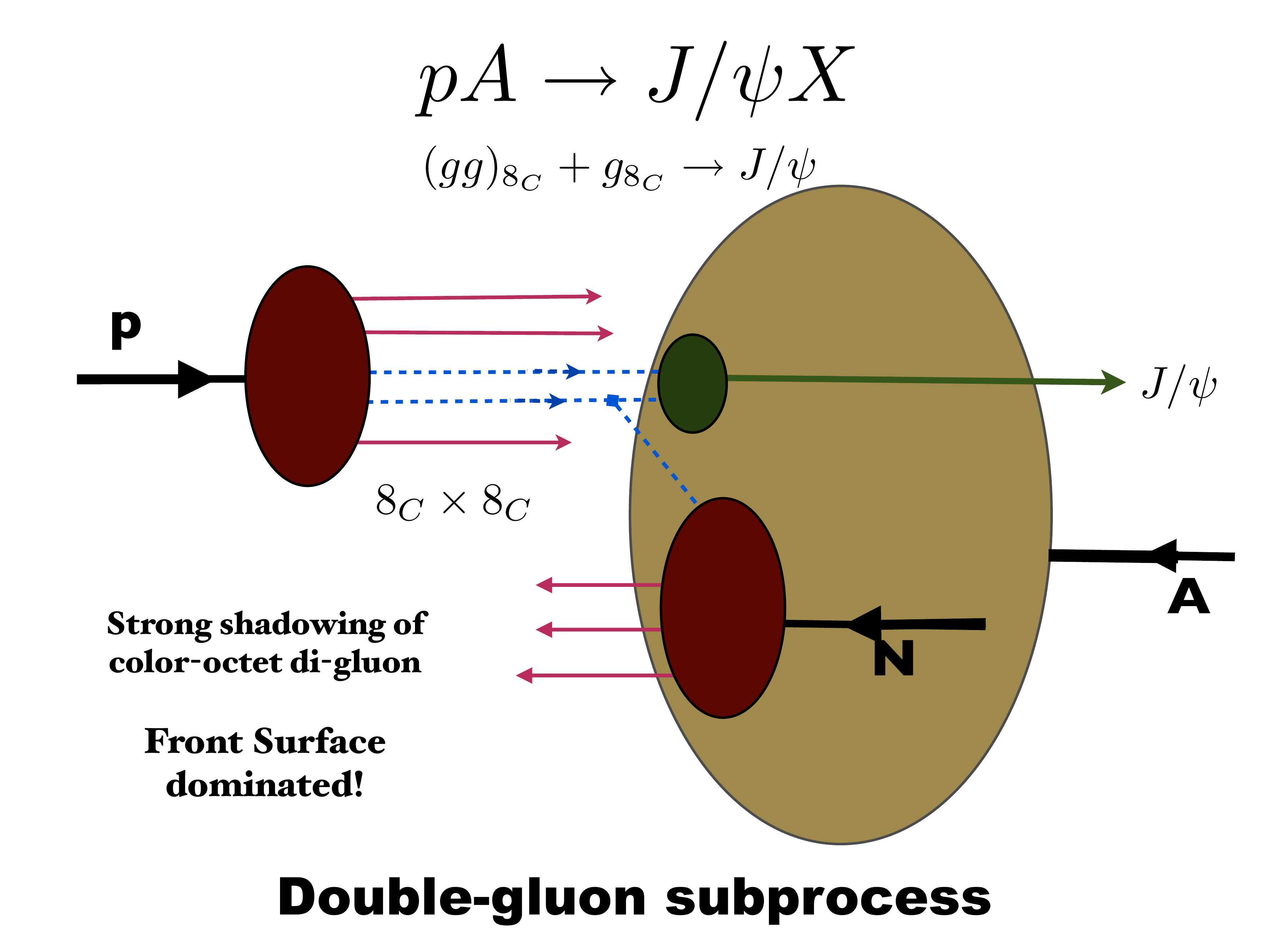}
\end{center}
\caption{Di-gluon mechanism for $J/\psi$ forward production.} 
\label{Digluon}  
\end{figure}

HuaXing Zhu and I have postulated an alternative mechanism:   We assume that the basic QCD mechanism for $J/\psi$ production at small transverse momentum  is $[gg] g \to J/\psi$ where the $[gg]$ is a color-octet di-gluon from the proton.  See fig. \ref{Digluon}.  The propagating color-octet di-gluon has a large interaction cross section, and it thus interacts primarily at the nucleus front surface, giving a production cross section  $\sigma(p A \to J/\psi X) \propto A^{2/3}.$ It should be noted that since $g(x,Q^2)$ falls rapidly, two gluons in the di-gluon,  each with $x \sim 0.01$, have a higher probability than a single gluon with $x \sim 0.02$.  The  di-gluon mechanism is expected to diminish in strength at increasing $p_T.$ 

The di-gluon subprocess is the color-octet analog of the color-singlet  two-gluon exchange mechanism~\cite{Brodsky:2002ue} underlying diffractive processes like $\ell p \to \ell p X$.    

The di-gluon multiparton subprocess is  analogous to the higher-twist subprocess $[q\bar] q q \to \gamma^* q $ which  dominates the 
$\pi N \to  \ell \bar \ell X$  Drell-Yan reaction at  high $x_F$, accounting for the observed dramatic change from transverse to longitudinal virtual photon polarization~\cite{Berger:1979du}.  Similarly,  multiparton ``direct" subprocesses can account~\cite{Arleo:2009ch} for the observed anomalous power-law fall-off of high $p_T$ inclusive hadron production cross sections ${ d \sigma\over d^3p/E}(p p \to h X)$ at fixed $x_T = 2{p_T\over \sqrt s} $ and fixed $\theta_{CM}$.

The $ pA \to J/\psi X$ cross sections measured in fixed-target experiments at CERN and FermiLab at high $x_F$ also show strong nuclear suppression at high $x_F$.  The ratio of the nuclear and proton target cross sections has the form $A^{\alpha(x_F)}$, where $x_F$ is Feynman fractional longitudinal momentum of the $J/\psi$. At small $x_F$, $\alpha(x_F)$  is slightly smaller than one, but at $x_F \sim 1$, it decreases to $\alpha=2/3$. These results  are again surprising since (1) the $\alpha= 2/3$ is characteristic of a strongly interacting hadron, not a small-size quarkonium state; and (2) the functional dependence   
$A^{\alpha(x_F)}$ contradicts  pQCD factorization~\cite{Hoyer:1990us}.

\begin{figure}
 \begin{center}
\includegraphics[height=10cm,width=15cm]{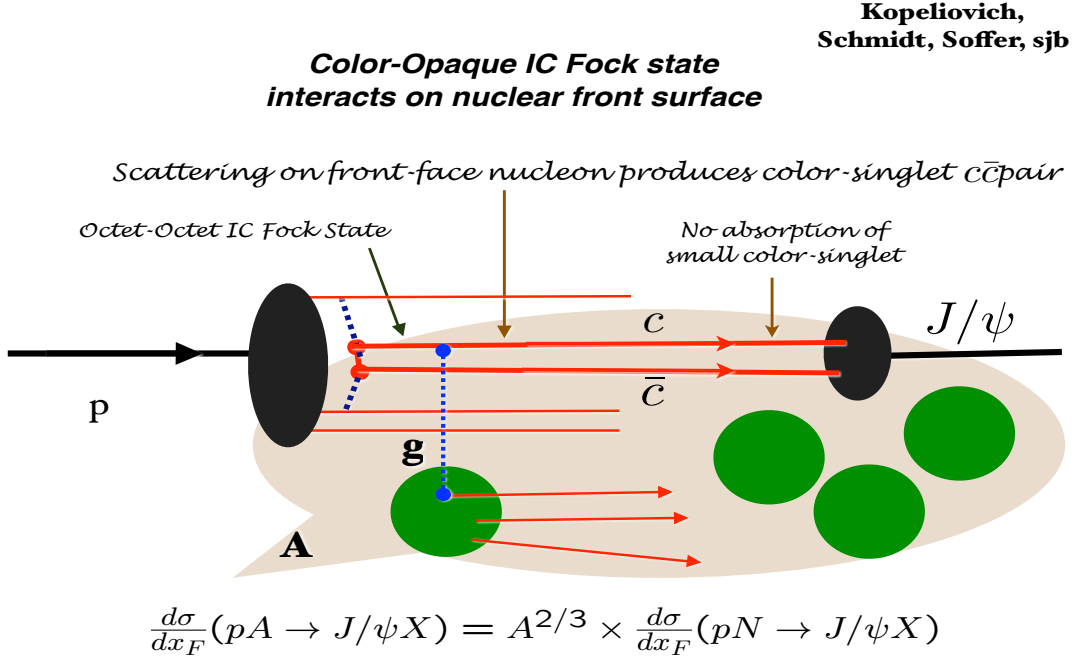}
\end{center}
\caption{Intrinsic charm model~\cite{Brodsky:2006wb,Brodsky:2007yz} for the nuclear dependence of J/psi hadroproduction at high $x_F$.} 
\label{AdepIC}  
\end{figure}

The observed nuclear suppression, in combination with the anomalously nearly flat cross section at high $x_F$, points to a QCD mechanism based on
the intrinsic charm Fock state~\cite{Brodsky:2007yz}.
QCD predicts that the color-configuration of the heavy quark pair $Q \bar Q$ in the intrinsic five-quark Fock state  is primarily a color-octet.
The intrinsic heavy quark Fock state of the proton: $|(uud)_{8_C} (c \bar c)_{8_C} \rangle$ thus interacts primarily with the $A^{2/3} $ nucleons at the front surface because of the large color-dipole moment of the color-octet $c \bar c$.    The $c \bar c$ color octet thus interacts primarily on a front-surface nucleon, changes to a color singlet, and then propagates through the nucleus as a $J/\psi$ at high $x_F$.   See fig. \ref{AdepIC}.

\section{Conclusions}

The phenomenological disagreements with conventional pQCD predictions discussed in this  contribution are not necessarily due to new physics beyond the standard model;  instead they may point to features of QCD itself, such as  intrinsic heavy quarks at high $x$  and the need to determine the appropriate renormalization scales.   I have outlined a number of novel physics consequences of intrinsic heavy quarks, such as the hadroproduction of the Higgs and exotic heavy quark states, both at high $x_F$ and at threshold.   Multi-parton subprocesses, such as di-gluon initiated reactions, can also play an important role.   I have discussed how AdS/QCD and light-front holography  provide a new analytic approach to the QCD confinement potential, hadron spectroscopy, hadron  dynamics, and the origin of the QCD mass scale.  I have also emphasized that the renormalization scale ambiguity can be consistently eliminated at finite orders in pQCD using the scheme-independent PMC procedure, thus eliminating an unnecessary source of theoretical systematic error.

\section{Acknowledgements}
Presented at the XV
International Conference on Hadron Spectroscopy, November 4-8, 2013, Nara, Japan.
I thank my collaborators for many helpful conversations.
This research was supported by the Department of Energy  contract DE--AC02--76SF00515.  
SLAC-PUB-15883.

\end{document}